\documentclass[american,prl,twocolumn]{revtex4-1}
\usepackage[T1]{fontenc}
\usepackage[latin9]{inputenc}
\usepackage{amsmath}
\usepackage{amssymb}
\usepackage{graphicx}

\makeatletter

\providecommand{\tabularnewline}{\\}

\makeatother

\usepackage{babel}
\begin{document}
\title{Spiral chains in wavenumber space of two dimensional turbulence}
\author{Ö. D. Gürcan$^{1}$, Shaokang Xu$^{1,3}$, P. Morel$^{1,2}$}
\affiliation{$^{1}$ Laboratoire de Physique des Plasmas, CNRS, Ecole Polytechnique,
Sorbonne Université, Université Paris-Saclay, Observatoire de Paris,
F-91120 Palaiseau, France~\\
$^{2}$ Département de Physique, Université Paris-Sud, Orsay, France~\\
$^{3}$ Peking University, School of Physics, Beijing, China}
\begin{abstract}
Self-similar, fractal nature of turbulence is discussed in the context
of two dimensional turbulence, by considering the fractal structure
of the wave-number domain using spirals. In loose analogy with phyllotaxis
in plants, each step of the cascade can be represented by a rotation
and a scaling of the interacting triad. Using a constant divergence
angle and a constant scaling factor, one obtains a family of such
fractals depending on the distance of interactions. Scaling factors
in such sequences are given by the square roots of known ratios such
as the plastic ratio, the supergolden ratio or some small Pisot numbers.
While spiral chains can represent mono-fractal models of self-similar
cascade, which can span a very large range in wave-number domain with
good angular coverage, it is also possible that spiral chains or chains
of consecutive triads play an important role in the cascade. As numerical
models, the spiral chain models based on decimated Fourier coefficients
have problems such as the dual cascade being overwhelmed by statistical
chain equipartition due to almost stochastic evolution of the complex
phases. A generic spiral chain model based on evolution of energy
is proposed, which is shown to recover the dual cascade behavior in
two-dimensional turbulence.
\end{abstract}
\maketitle

\section{Introduction}

Spiral patterns emerge in many nonlinear problems in nature, from
galaxy formation to crystal growth, from plants to animals and from
atmospheric cyclones to small scale turbulence, they appear at very
different scales and in very different problems. They are a fundamental
element of phyllotaxis -the dynamical phenomenon of arrangement of
seeds or petals of a plant (sometimes in the form of flowers) as it
grows\citep{adler:97}. Mathematically, the particularity of the spiral
form is that it keeps certain quantities (such as the angle between
two consecutive elements) invariant as the structure is scaled and
rotated. This provides a natural self-similar framework with which
the some physical systems operate. One of the key aspect of phyllotaxis
is how a discrete structure that grows through iteration manages optimal
packing, leading to the observed fractal pattern\citep{douady:96,newell:08}.
Similar concepts apply to reaction-diffusion systems where spiral
patterns arise in a continuum of deformations\citep{crampin:02}.
Incidentally, spiral patterns also occur in turbulence\citep{lungren:82},
especially in two dimensions\citep{gilbert:88,bouchet:12}, mainly
as a result of self shearing of smaller scale structures by large
scale flows, and the resulting self-similarity of the turbulent flow,
where the structure remains the same as it scales and turns. In fact
the basic motion of scale and rotate (i.e. ``swirl''), associated
with a turbulent flow naturally implies a spiral-like pattern.

Spirals in wave-vector space are also potentially interesting for
the study of turbulent dynamics. Common sense suggests that nonlinear
interactions tend to scale and rotate real space structures, and hence
they would do the same to the wave-vectors as well. For instance,
if we have a particular direction of anisotropy, at a given scale,
nonlinearity tends to generate a ``next'' scale in the hierarchy,
which is anisotropic in a direction that is ``at a certain angle''
(maybe perpendicular) to the original direction of anisotropy. Thus,
when there is a large scale source of anisotropy, going towards smaller
scales the direction of anisotropy at each scale keeps changing, which
results in a virtually isotropic spectrum in statistical sense.

Energy (and enstrophy for two dimensions), gets transferred via triadic
interactions in turbulent flows\citep{frisch}. In general for a given
scale, there are many such triads that can transfer energy or enstrophy
in either directions to other scales. If, for some reason, one of
these triads is ``dominant'' -for example due to the fact that it
maximizes the interaction coefficient-, it is natural that this triad
will take more of the energy or enstrophy along. Then, at the next
scale the energy goes, the ``same triad'' (now rotated and scaled),
will likely win again for the same reason that it won at the first
scale, transferring the energy to the next one along a chain of such
dominant triads. It is unclear if the small differences among nearby
triads in terms of their capacity to transfer energy and enstrophy
justifies a reduction of the turbulent transfer to picture of transfer
along a single chain of scaled and rotated triads that arrange naturally
into a spiral. Nonetheless the picture of turbulent energy transfer
as taking place along chains of spirals (instead of the naive and
incorrect picture of a ``radial'' flux in $k$-space) that compete
with and couple to one another is instructive.

Various kinds of reduced models have been proposed in the past, in
order to study both the nonlinear cascade and the direction of anisotropy
in turbulent flows from shell models\citep{ohkitani:89,biferale:03},
to differential approximation models\citep{leith:67,lilly:89,lvov:06}
to closure based models\citep{kraichnan:59,orszag:70} to tree models\citep{aurell:94,aurell:97,waleffe:06}
to reduced wave-number representations\citep{grossmann:96}. Here
we propose a reduction of two dimensional turbulence based on spiral
chains, which are chains of wave-numbers that are obtained by scaling
and rotating a single triad such that the smaller wave-number of the
triad, after scaling and rotation (or after a few scalings and rotations),
becomes, first the middle wave-number and then the larger wave-number.
In principle a number of such spiral chains can be used, instead of
a single one, in order to span the $k$-space more completely.

The rest of the paper is organized as follows. In section \ref{sec:single_triad},
the problem of a single triad is revisited and the concept of triad
chains or consecutive triads by which the energy is transferred is
discussed. In section \ref{sec:Spiral-Chain-Models}, regular spiral
chain models for certain chains with relatively local interactions
are introduced. The general case of arbitrarily distant interactions
is also covered in this section where a list of possible values of
scaling factors and divergence angles are given in table \ref{tab:lm}.
Possible stationary solutions are discussed in Section \ref{subsec:soln1},
conservation of energy and enstrophy for spiral chains is formulated
in Section \ref{subsec:Energy-and-Enstrophy:} and zero flux solutions
are investigated in Section \ref{subsec:Zero-flux-solutions}. In
Section \ref{sec:The-model-for} a spiral chain model formulated for
chain energy $E_{n}$ is introduced. Re-interpreting this model as
a model for shell energy, with the assumption of isotropy, which allows
the interactions to be infinitesimally local, the continuum limit
is computed and found to be the usual differential approximation model
form for the two dimensional Euler turbulence in section \ref{subsec:Continuum-limit}.
A four spiral chain model with good angular covergae is introduced
in \ref{subsec:4-Spiral-Chain-Model}. Numerical results for a subset
of these spiral chain models are given in \ref{sec:Numerical-Results}.
Section \ref{sec:Conclusion} is conclusion.

\section{Dynamics of a Single Triad\label{sec:single_triad}}

Two dimensional turbulence, as represented by an equation of advection
of vorticity\citep{kraichnan:80}, or more generally, of potential
vorticity\citep{pedlosky-book:71} can be relevant as a simplified
limiting case of many physical problems from rotating turbulence in
laboratory experiments\citep{godeferd:15}, to geostrophic turbulence
in planetary atmospheres\citep{rhines:79}, to drift wave turbulence
in tokamak plasmas\citep{horton:99}.

Consider the two dimensional Euler equation
\begin{equation}
\partial_{t}\nabla^{2}\Phi+\hat{\mathbf{z}}\times\nabla\Phi\cdot\nabla\nabla^{2}\Phi=0\;\text{,}\label{eq:2d_euler}
\end{equation}
to which viscosity or hyper-viscosity can be added for dissipation
of energy and enstrophy. Its Fourier transform can be written in general
as
\[
\partial_{t}\Phi_{k}=\sum_{p+q=-k}\frac{\hat{\mathbf{z}}\times\mathbf{p}\cdot\mathbf{q}\left(q^{2}-p^{2}\right)}{k^{2}}\Phi_{p}^{*}\Phi_{q}^{*}
\]
with the convention that $\sum_{p+q=-k}$ represents a sum over $\mathbf{p}$
and $\mathbf{q}$ such that $\mathbf{k}+\mathbf{p}+\mathbf{q}=0$.
Now consider a single triad consisting of $\mathbf{k}$, $\mathbf{p}$
and $\mathbf{q}$ such that $k<p<q$. If $\eta\equiv\frac{\ln\left(q/k\right)}{\ln\left(p/k\right)}\in\mathbb{Q}$
(i.e. is rational) we can write $p=kg^{\ell}$ and $q=kg^{m}$ (i.e.
$\eta=m/\ell$). Obviously not all triangles satisfy the condition
$\eta\in\mathbb{Q}$. However there is usually an approximately equivalent
triangle from a physics or numerics perspective, which does. If one
is restricted to low order rationals for $\eta$, it is only a particular
class of triangles, which can be represented as $p=kg^{\ell}$ and
$q=kg^{m}$ with $\ell$ and $m$ integers and $g>1$ (i.e. $g\in\mathbb{R}$).

For those triangles, we can write the interaction as:
\[
\partial_{t}\Phi_{k}=k^{2}\sin\alpha_{qp}g^{m+\ell}\left(g^{2m}-g^{2\ell}\right)\Phi_{p}^{*}\Phi_{q}^{*}
\]
\[
\partial_{t}\Phi_{p}=k^{2}\sin\alpha_{qp}g^{m-\ell}\left(1-g^{2m}\right)\Phi_{q}^{*}\Phi_{k}^{*}
\]
\[
\partial_{t}\Phi_{q}=k^{2}\sin\alpha_{qp}g^{\ell-m}\left(g^{2\ell}-1\right)\Phi_{k}^{*}\Phi_{p}^{*}
\]
where we have used $\left(\hat{\mathbf{z}}\times\hat{\mathbf{p}}\cdot\hat{\mathbf{q}}\right)=\sin\alpha_{qp}=\sin\left(\theta_{q}-\theta_{p}\right)$.
Since $g>1$ and the middle leg of the triad (i.e. $p$) is unstable
as long as $m>\ell$ (which we have assumed by assuming $q>p$) and
gives its energy to the other two wave-numbers.

The energy evolves according to
\[
\partial_{t}E_{k}=\left(g^{2m}-g^{2\ell}\right)t_{kpq}
\]
\[
\partial_{t}E_{p}=\left(1-g^{2m}\right)t_{kpq}
\]
\[
\partial_{t}E_{q}=\left(g^{2\ell}-1\right)t_{kpq}
\]
where
\[
t_{kpq}=g^{m+\ell}k^{4}\sin\alpha_{qp}\Phi_{p}^{*}\Phi_{q}^{*}\Phi_{k}^{*}
\]
It is easy to see that the total energy of the triad is conserved.
Following the reasoning discussed in Ref. \citep{depietro:15}, the
instability assumption implies $\overline{t}_{kpq}>0$ since $\overline{E}_{p}$
should decrease in time, where the overbar implies statistical ensemble
average, which can be replaced by time average in most cases.

The energy that is transferred from $p$ to $k$ is $g^{2m}t_{kpq}$,
while the energy that is transfered from $q$ to $p$ is simply $t_{kpq}$.
On the other hand there is energy that is transfered from $k$ to
$q$ (from the smallest to the largest wave-number), which is $g^{2\ell}t_{kpq}$.
Since $g^{2m}>g^{2\ell}$, $E_{k}$ gets more energy than it looses.
However since $g^{2\ell}>1$, $E_{q}$ also gets more energy than
it looses. This means the energy is transfered from the middle wave-number
to the larger and smaller wave-numbers. If the sign of $t_{kpq}$
changes, then the flow will be towards the middle wave-number, in
fact the system will naturally undergo such oscillations as the energy
of the middle wave-number gets depleted.

\subsection{Consecutive triads:}

Imagine the triad $\mathbf{k}$, $\mathbf{p}$, $\mathbf{q}$ discussed
above. If we scale it by $g^{-\ell}$ and rotate by $-\theta_{p}$,
we obtain a second triad where $\mathbf{k}$ becomes the middle wave-number
instead of the smallest one (we call the other two wave-numbers as
$p'$ and $q'$ with $p'<k<q'$). and if we scale it by $g^{-m}$,
and rotate by $-\theta_{q}$, $\mathbf{k}$ becomes the largest wave-number
(with $p''$ and $q''$ such that $p''<q''<k$). Note that $p'=kg^{-\ell}$,
$q'=kg^{m-\ell}$, $p''=kg^{-m}$, $q''=kg^{\ell-m}$. By defining
$k\rightarrow k_{n}$, and assuming that those three triads exist,
we can write the evolution equation for $\Phi_{k_{n}}\rightarrow\Phi_{n}$
as
\begin{align}
\partial_{t}\Phi_{n}=k_{n}^{2}\sin\alpha_{qp}\bigg[ & g^{m+\ell}\left(g^{2m}-g^{2\ell}\right)\Phi_{n+\ell}^{*}\Phi_{n+m}^{*}\nonumber \\
 & +g^{m-3\ell}\left(1-g^{2m}\right)\Phi_{n-\ell+m}^{*}\Phi_{n-\ell}^{*}\nonumber \\
 & +g^{\ell-3m}\left(g^{2\ell}-1\right)\Phi_{n-m}^{*}\Phi_{n-m+\ell}^{*}\bigg]\;\text{.}\label{eq:goy_gen}
\end{align}
The three terms on the right hand side of (\ref{eq:goy_gen}) are
the contributions from $\left(p,q\right)$, $\left(p',q'\right)$
and $\left(p'',q''\right)$ respectively or to the three triangles
from the largest to the smallest. Note that for a given triangle shape,
the three terms in (\ref{eq:goy_gen}) appear naturally representing
the three different size triangles (but of the same shape), where
$\mathbf{k}$ play the role of the smallest, the middle and the largest
wave-numbers consecutively. In fact one can also imagine adding a
sum over different shapes of triangles in order to provide a complete
description.

If we call the triangles from the smallest to the largest as $\triangle_{1}$,
$\triangle_{2}$ and $\triangle_{3}$ respectively, we obtain $\triangle_{2}$
by scaling $\triangle_{1}$ by $g^{m-\ell}$ and rotating it by $\alpha_{qp}=\theta_{q}-\theta_{p}$,
and $\triangle_{3}$, by scaling $\triangle_{2}$ by $g^{\ell}$ and
rotating it by $\theta_{p}$. Obviously we can repeat the procedure
of rotating and scaling in order to cover a whole range of $k$ vectors
in the wave-number domain. However while the scaling is regular (i.e.
we can define a $k_{n}=k_{0}g^{n}$ such that scaled wave-numbers
always have the form $k_{n}$ with $n\in\mathbb{Z}$), in general
the angles are not perfectly regular.

\begin{figure}
\begin{centering}
\includegraphics[width=0.8\columnwidth]{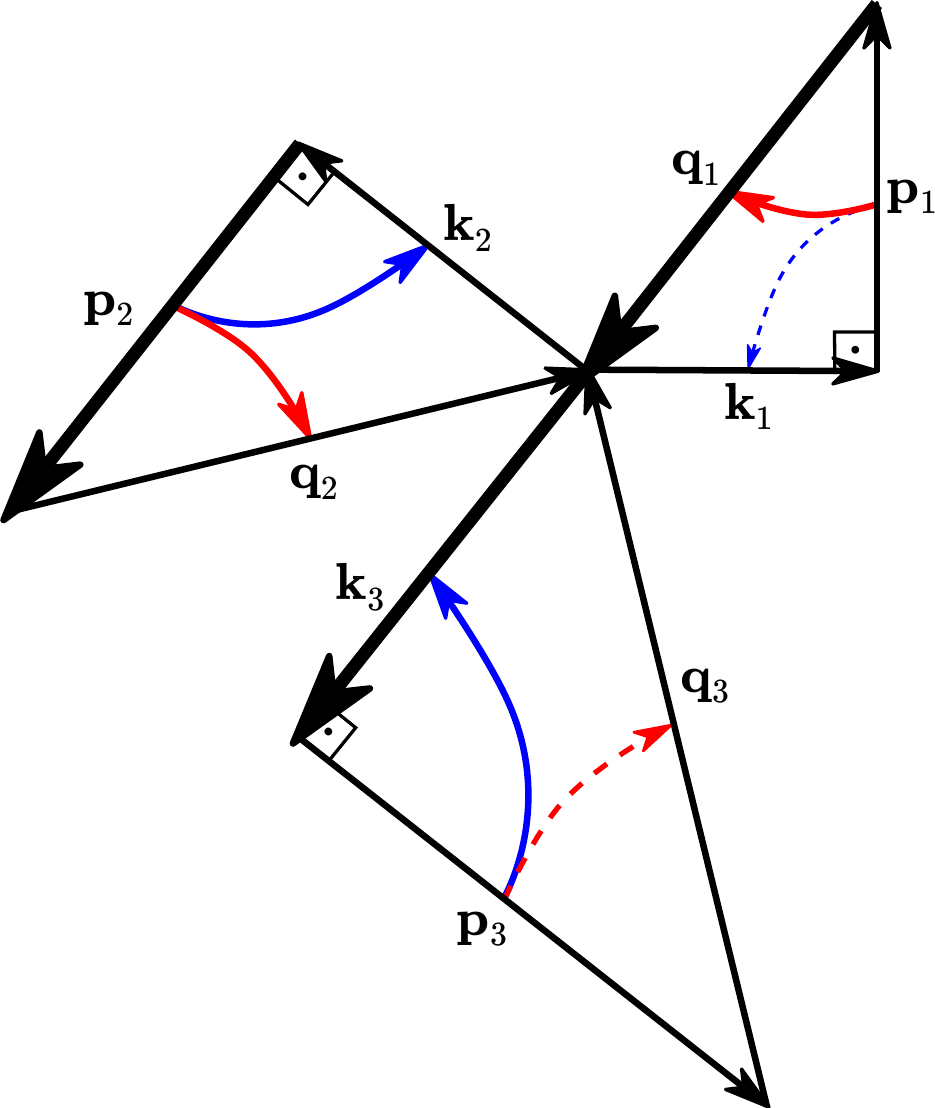}
\par\end{centering}
\caption{\label{fig:three_triangles}The triad $\triangle_{1}$ defined as
$\ell=1$, $m=2$, $g=\sqrt{\varphi}$. Scaling $\triangle_{1}$ by
$g$ and rotating by $\pi/2$, we obtain $\triangle_{2}$. Scaling
$\triangle_{2}$ by $g$ and rotating by $\alpha_{qp}=141.83^{0}$,
we obtain $\triangle_{3}$. Note that the three triads share the common
wave-vector $\mathbf{q}_{1}=\mathbf{p}_{2}=\mathbf{k}_{3}$, which
we can call $\mathbf{k}_{n}$. The energy inverse cascades via $\mathbf{p}_{3}\rightarrow\mathbf{k}_{n}\rightarrow\mathbf{k}_{2}$
(blue arrows) while enstrophy forward cascades via $\mathbf{p}_{1}\rightarrow\mathbf{k}_{n}\rightarrow\mathbf{q}_{2}$
(red arrows).}

\end{figure}

Consider for example the triangle with $g=\sqrt{\varphi}$ where $\varphi=\left(1+\sqrt{5}\right)/2$
is the golden ratio so that $k=1$, $p=g$ and $q=g^{2}$. The angle
between $k$ and $p$ is a right angle (since $\sqrt{1+g^{2}}=g^{2}$
with $g=\sqrt{\varphi}$), while the one between $p$ and $q$ can
be computed from the law of cosines as $\cos\alpha_{qp}=\frac{1-p^{2}-q^{2}}{2qp}=\frac{1-g^{2}-g^{4}}{2g^{3}}$,
which gives an angle about $\alpha_{pq}=141.83^{0}$ (note that $\alpha_{pq}$
is the angle between the two vectors, which is $\pi$ minus the angle
between the two edges of the triangle). This corresponds to the triangle
defined by $\ell=1$, $m=2$ and $g=\sqrt{\varphi}$. Scaling this
triangle $\triangle_{1}$ by $g$ and rotating by $\pi/2$, we obtain
triangle $\triangle_{2}$, scaling $\triangle_{2}$ by $g$ and rotating
by $141.83^{0}$ we obtain $\triangle_{3}$. We can can construct
a chain of such triads that are connected to one another by the common
wave-number as shown in figure \ref{fig:three_triangles}, for which
the equation of motion will still be (\ref{eq:goy_gen}). However
the grid that is gnereated by the triad chain is, in general, irregular.

However, it is obvious from this emerging picture that if we had $\alpha_{qp}=m\alpha_{pk}$
where $m$ is some integer, we could write the whole thing as a regular
spiral, with $k_{n}=k_{0}g^{n}$ and $\theta_{n}=n\alpha$. It is
also obvious that the class of triangles that would result in such
a regular spiral, are a very special class: Each wave-number involved
in such a system is a rotated and scaled version of the wave-number
before it in a regular fashion.

\section{Spiral Chain Models\label{sec:Spiral-Chain-Models}}

Let us introduce the symbol $\mathcal{C}_{\ell m}^{s_{\ell}s_{m}}$
to refer to a basic spiral chain consisting of the triad $\mathbf{k}_{n}+s_{\ell}\mathbf{k}_{n+\ell}+s_{m}\mathbf{k}_{n+m}=0$,
where $k_{n}=k_{0}g^{n}$ and $\theta_{n}=\alpha n$ {[}or using the
equivalence between two dimensional vectors and complex numbers, $k_{n}^{c}=k_{0}\left(ge^{i\alpha}\right)^{n}$
with $\mathbf{k}_{n}=\text{Re}\left(k_{n}^{c}\right)\hat{\boldsymbol{x}}+\text{Im}\left(k_{n}^{c}\right)$$\hat{\boldsymbol{y}}${]}.
Note that $g$ and $\alpha$ follows from $\ell$, $m$, $s_{\ell}$
and $s_{m}$, and therefore need not be stated explicitly. Here $s_{\ell}$
and $s_{m}$ are the signs in front of the wave-numbers in order to
satisfiy the triad condition.

Considering $\ell=2$, $m=3$ in (\ref{eq:goy_gen}), with $\theta_{n}=n\alpha$,
so that $\alpha_{pk}=2\alpha$, $\alpha_{qp}=\alpha$ and $\alpha_{qk}=3\alpha$,
and all possible interaction forms (i.e. $\mathbf{k}\pm\mathbf{p}\pm\mathbf{q}=0$),
we find that the law of cosines for the different cases give
\[
\cos\alpha_{pk}=\pm\left(\frac{q^{2}-k^{2}-p^{2}}{2kp}\right)=\pm\left(\frac{g^{6}-g^{4}-1}{2g^{2}}\right)=\cos2\alpha
\]
\[
\cos\alpha_{qp}=\pm\left(\frac{k^{2}-p^{2}-q^{2}}{2pq}\right)=\pm\left(\frac{1-g^{4}-g^{6}}{2g^{5}}\right)=\cos\alpha
\]
\[
\cos\alpha_{qk}=\pm\left(\frac{p^{2}-q^{2}-k^{2}}{2qk}\right)=\pm\left(\frac{g^{4}-1-g^{6}}{2g^{3}}\right)=\cos3\alpha
\]
where the sign $\pm$ corresponds to the relative sign of the two
corresponding wave-numbers (e.g. $p$ and $k$ for $\alpha_{pk}$)
in the expression $\mathbf{k}\pm\mathbf{p}\pm\mathbf{q}=0$. We can
obtain two polynomial relations for $g$ using the trigonometric relations
$\cos2\alpha=2\cos^{2}\alpha-1$ and $\cos3\alpha=\cos\alpha\left(4\cos^{2}\alpha-3\right)$.
Both of these can be solved for the cases $\mathbf{k}-\mathbf{p}+\mathbf{q}=0$
and $\mathbf{k}-\mathbf{p}-\mathbf{q}=0$ with $g\approx1.15096$
and an angle $\alpha=\arccos\left(-g^{3}/2\right)$ for the case $\mathbf{k}-\mathbf{p}-\mathbf{q}=0$
or $\alpha=\pi-\arccos\left(-g^{3}/2\right)$ for the case $\mathbf{k}-\mathbf{p}+\mathbf{q}=0$.
Note that the actual positive root ($g>1$ ) of the polynomial equation
is $g=\sqrt{\rho}$ where $\rho$ is the plastic number, whose exact
value can be written as:
\[
\rho=\left(\frac{1}{2}\right)^{1/3}\left[\left(1-\sqrt{\frac{23}{27}}\right)^{1/3}+\left(1+\sqrt{\frac{23}{27}}\right)^{1/3}\right]
\]
\begin{figure}
\includegraphics[width=1\columnwidth]{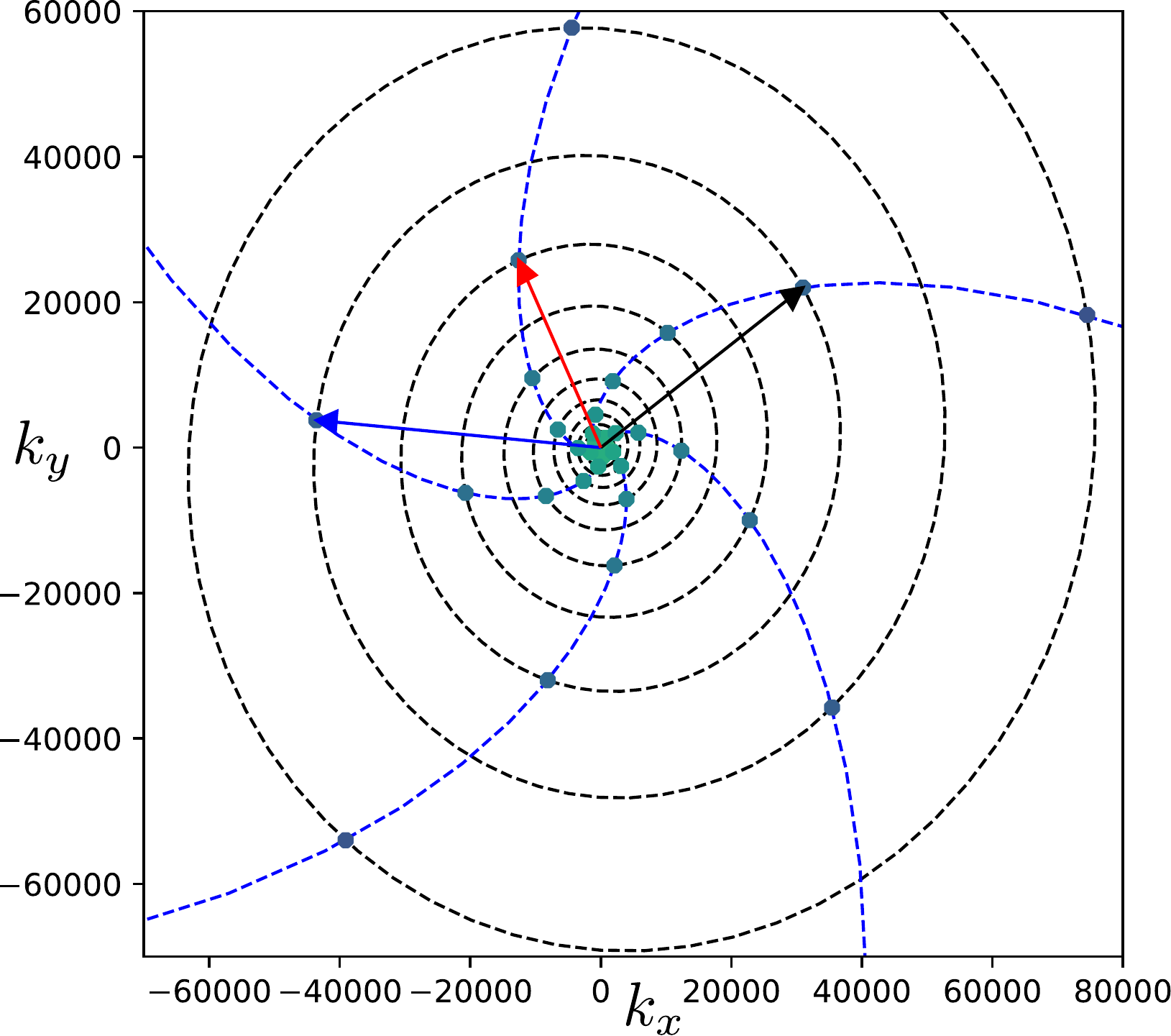}

\caption{\label{fig:spiral}The spiral chain $\ell=2$, $m=3$ with $g=\sqrt{\rho}$.
The counter clockwise primary spiral chain is shown in black dashed
lines while the clockwise secondary spirals are shown in blue dashed
lines. Note that as the energy travels along the primary chain, it
gets exchanged between the $5$ secondary chains. Finally an interacting
triad with $\mathbf{k}=\mathbf{k}_{n}$ (black arrow), $\mathbf{p}=\mathbf{k}_{n-2}$
(red arrow) and $\mathbf{q}=\mathbf{k}_{n+1}$ (blue arrow) is shown
(i.e. $\mathbf{k}+\mathbf{q}-\mathbf{p}=0$).}
\end{figure}

\subsubsection{Chain $\mathcal{C}_{2,3}^{-,-}$}

For this basic chain, which can be denoted by $\mathcal{C}_{2,3}^{-,-}$
a basic evolution equation can be written as follows:
\begin{align}
\partial_{t}\Phi_{n} & =k_{n}^{2}\sin\alpha\bigg[g^{-7}\left(g^{4}-1\right)\Phi_{n-3}\Phi_{n-1}^{*}\nonumber \\
 & -g^{-3}\left(g^{6}-1\right)\Phi_{n-2}\Phi_{n+1}^{*}\nonumber \\
 & +g^{9}\left(g^{2}-1\right)\Phi_{n+2}\Phi_{n+3}\bigg]+P_{n}-D_{n}\label{eq:spi1}
\end{align}
with $\Phi_{n}=\hat{\Phi}\left(\mathbf{k}_{n}\right)$ as the Fourier
coefficient of $\Phi$, with the wavevector \textbf{$\mathbf{k}_{n}=k_{n}\left(\cos\alpha_{n},\sin\alpha_{n}\right)$},
where $k_{n}=k_{0}g^{n}$ and $\alpha_{n}=\alpha n$, $g=\sqrt{\rho}$
being the logarithmic scaling factor and $\alpha=\arccos\left(-g^{3}/2\right)$,
being the divergence angle. $P_{n}$ and $D_{n}$ are energy injection
and dissipation respectively (i.e. $D_{n}=\nu k_{n}^{2}\Phi_{n}$
for a usual kinematic viscosity and $P_{n}=\gamma_{n}\Phi_{n}$ for
an internal instability drive).

Note that using the relations $g^{6}-1=g^{2}$, $g^{4}-1=g^{-2}$
and $g^{2}-1=g^{-8}$, possible due to the choice $g=\sqrt{\rho}$,
we can write (\ref{eq:spi1}) also as:
\begin{align}
\partial_{t}\Phi_{n} & =k_{n}^{2}\sin\alpha\bigg[g^{-9}\Phi_{n-3}\Phi_{n-1}^{*}-g^{-1}\Phi_{n-2}\Phi_{n+1}^{*}\nonumber \\
 & +g\Phi_{n+2}\Phi_{n+3}\bigg]+P_{n}-D_{n}\Phi_{n}\label{eq:spi2}
\end{align}
While (\ref{eq:spi2}) conserves energy and enstrophy for $g=\sqrt{\rho}$,
(\ref{eq:spi1}) does so for arbitrary $g$, which makes it somewhat
more useful even though the two equations are identical for the given
value of $g$.

\subsubsection{Chain $\mathcal{C}_{2,3}^{-,+}$}

It is clear that there are many similar chains, such as the one with
$\alpha=\pi-\arccos\left(-g^{3}/2\right)=\arccos\left(g^{3}/2\right)$,
which gives a similar model, but with a different conjugation structure:
\begin{align}
\partial_{t}\Phi_{n} & =k_{n}^{2}\sin\alpha\bigg[g^{-7}\left(g^{4}-1\right)\Phi_{n-3}^{*}\Phi_{n-1}\nonumber \\
 & -g^{-3}\left(g^{6}-1\right)\Phi_{n+1}\Phi_{n-2}\nonumber \\
 & +g^{5}\left(g^{6}-g^{4}\right)\Phi_{n+2}\Phi_{n+3}^{*}\bigg]+P_{n}-D_{n}\label{eq:spi1b}
\end{align}
and a different sampling of wave-vector directions.

\subsubsection{Chain $\mathcal{C}_{-1,2}^{+,+}$ (or $\mathcal{C}_{1,3}^{+,+}$)}

We can obtain another chain by choosing $\ell=-1$, $m=2$, which
gives $\alpha_{pk}=-\alpha$, $\alpha_{qp}=3\alpha$ and $\alpha_{qk}=2\alpha$.
Using the law of cosines and the relations between $\cos\alpha$,
$\cos2\alpha$ and $\cos3\alpha$, we obtain $g\approx1.21061$, or
$g=\sqrt{\psi}$ where:
\[
\psi=\frac{1}{3}\left[1+\frac{1}{2^{1/3}}\left(\left(29+3\sqrt{93}\right)^{1/3}+\left(29-3\sqrt{93}\right)^{1/3}\right)\right]
\]
is the so-called super golden ratio, and $\alpha=\arccos\left(g^{-3}/2\right)$
for the form $\mathbf{k}+\mathbf{p}+\mathbf{q}=0$, and thus an evolution
equation of the form:
\begin{align}
\partial_{t}\Phi_{n} & =k_{n}^{2}\sin\alpha\bigg[g^{-11}\left(g^{2}-1\right)\Phi_{n-2}^{*}\Phi_{n-3}^{*}\nonumber \\
 & -g^{-3}\left(g^{6}-1\right)\Phi_{n-1}^{*}\Phi_{n+2}^{*}\nonumber \\
 & +g^{3}\left(g^{4}-1\right)\Phi_{n+3}^{*}\Phi_{n+1}^{*}\bigg]+P_{n}-D_{n}\label{eq:spi2a}
\end{align}
where we have used the fact that for this particular value of $\alpha$,
we have $\sin3\alpha=-g^{-2}\sin\alpha$.

\subsubsection{Chain $\mathcal{C}_{-1,2}^{-,+}$ (or $\mathcal{C}_{1,3}^{-,-}$)}

A similar case to chain $\mathcal{C}_{-1,2}^{+,+}$ exists with $g=\sqrt{\psi}$
and $\alpha=\arccos\left(-g^{-3}/2\right)=\pi-\arccos\left(g^{-3}/2\right)$,
which corresponds to $\mathbf{k}+\mathbf{q}-\mathbf{p}=0$ and the
evolution equation of the form:
\begin{align*}
\partial_{t}\Phi_{n} & =k_{n}^{2}\sin\alpha\bigg[g^{-11}\left(g^{2}-1\right)\Phi_{n-2}^{*}\Phi_{n-3}\\
 & -g^{-3}\left(g^{6}-1\right)\Phi_{n-1}\Phi_{n+2}^{*}\\
 & +g^{3}\left(g^{4}-1\right)\Phi_{n+3}\Phi_{n+1}\bigg]+P_{n}-D_{n}
\end{align*}
The chain denoted by $\ell=1$, $m=3$ corresponds to the same chain
as the one denoted by $\ell=-1$, $m=2$. (since we can obtain one
from the other by exchanging $k$ and $p$). This means we can write
$\mathcal{C}_{-1,2}^{+,+}=\mathcal{C}_{1,3}^{+,+}$ and $\mathcal{C}_{-1,2}^{-,+}=\mathcal{C}_{1,3}^{-,-}$
or in general $\mathcal{C}_{\ell,m}^{s_{\ell},s_{m}}=\mathcal{C}_{-\ell,m-\ell}^{s_{\ell},s_{m}*s_{\ell}}$.
This means that it is sufficient to consider the case $m>\ell>0$.

\subsubsection{Chains $\mathcal{C}_{2,3}^{-,-}+\mathcal{C}_{1,5}^{-,-}$}

Remarkably, the case $\ell=1$ and $m=5$ gives $g=\sqrt{\rho}$ and
$\alpha=\arccos\left(-g^{3}/2\right)$ exactly as in the case $\ell=2$
and $m=3$. This means that in fact these two spiral chains are inseparable
since a choice of $g$ and $\alpha$, will lead to an evolution equation
of the form:

\begin{align}
\partial_{t} & \Phi_{n}=k_{n}^{2}\sin\alpha\bigg[-g^{-19}\left(g^{2}-1\right)\Phi_{n-5}\Phi_{n-4}^{*}\nonumber \\
 & +g^{-7}\left(g^{4}-1\right)\Phi_{n-3}\Phi_{n-1}^{*}-g^{-3}\left(g^{6}-1\right)\Phi_{n-2}\Phi_{n+1}^{*}\nonumber \\
 & +g^{-3}\left(g^{10}-1\right)\Phi_{n-1}\Phi_{n+4}^{*}-g^{3}\left(g^{8}-1\right)\Phi_{n+1}\Phi_{n+5}\nonumber \\
 & +g^{9}\left(g^{2}-1\right)\Phi_{n+2}\Phi_{n+3}\bigg]+P_{n}-D_{n}\;\text{,}\label{eq:spi_full_pre}
\end{align}
It is easy to show that these are in fact \emph{all the interactions}
that take place among the points of this particular spiral (i.e. defined
by $g$ and $\alpha$). Similarly there is another double chain of
the form $\mathcal{C}_{2,3}^{-,+}+\mathcal{C}_{1,5}^{+,+}$ as well.

\subsubsection{Supplementary chains}

Consider the two chains represented by $\mathcal{C}_{1,3}^{+,+}$
and $\mathcal{C}_{1,3}^{-,-}$ discussed above. The two chains have
the same $g$'s but supplementary angles. This means that while the
$++$ chain has the angles $\theta_{n}=n\alpha$, the supplementary
chain has the angles $\theta_{n}=n\left(\pi-\alpha\right)$. However
since both $\Phi_{n}$ and $\Phi_{n}^{*}$ are considered for a given
$\mathbf{k}_{n}$, adding or substracting $\pi$ to an angle is equivalent
to taking the complex conjugate or replacing $\mathbf{k}_{n}\rightarrow-\mathbf{k}_{n}$.
Therefore we can instead use $\theta_{n}=-n\alpha$, and note that
it corresponds to the spiral that rotates in the opposite direction
to the original spiral. But with $\mathbf{k}_{n}+\mathbf{k}_{n+1}+\mathbf{k}_{n+3}=0$,
since the signs of $k_{n\pm\ell}$ for odd $\ell$ change direction.

\subsubsection{Other Chains:}

If we consider other $\ell$ and $m$ values, it is clear that $\ell=4$,
$m=6$ gives $g_{4,6}=\left(g_{2,3}\right)^{1/2}$ and $\alpha_{4,6}=\alpha_{2,3}/2$
etc. These are not unique chains but simply the same chains that are
repeated twice {[}or $n$ times to get $g_{2n,3n}=\left(g_{2,3}\right)^{1/n}$,
and $\alpha_{2n,3n}=\alpha_{2,3}/n${]}. In contrast, for a unique
chain, we have to compute $g$ and $\alpha$. In general, for any
$\ell$ and $m$ such that $\mathbf{k}_{n}+s_{\ell}\mathbf{k}_{n+\ell}+s_{m}\mathbf{k}_{n+m}=0$,
we can write 
\[
\cos\ell\alpha=s_{\ell}\frac{\left(g^{2m}-g^{2\ell}-1\right)}{2g^{\ell}}
\]
\[
\cos m\alpha=s_{m}\frac{\left(g^{2\ell}-g^{2m}-1\right)}{2g^{m}}
\]
\[
\cos\left(m-\ell\right)\alpha=s_{m}s_{\ell}\frac{\left(1-g^{2m}-g^{2\ell}\right)}{2g^{\left(m+\ell\right)}}\;\text{.}
\]
Consistency requires that:
\begin{align}
 & \frac{1}{\ell}\arccos\left[s_{\ell}\frac{\left(g^{2m}-g^{2\ell}-1\right)}{2g^{\ell}}\right]\nonumber \\
 & =\frac{1}{m}\arccos\left[s_{m}\frac{\left(g^{2\ell}-g^{2m}-1\right)}{2g^{m}}\right]\nonumber \\
 & =\frac{1}{m-\ell}\arccos\left[s_{\ell}s_{m}\frac{\left(1-g^{2m}-g^{2\ell}\right)}{2g^{\left(m+\ell\right)}}\right]\label{eq:geq}
\end{align}
where the arccos function is considered as multi-valued. These equations
can be solved numerically in order to obtain spiral chains for any
$\ell$ and $m$ values. In general for a given $\ell$ and $m$,
one may have multiple solutions of (\ref{eq:geq}) because of the
multivaluedness of the arccosine functions. Note that the combination
of $s_{\ell}$ and $s_{m}$ and $g$ define a unique angle $\alpha$.
See table \ref{tab:lm} for the list of all possible chains up to
$m=9$. Note that for each chain that is represented in table \ref{tab:lm},
there is also the supplementary chain with $\alpha'=\pi-\alpha$ and
$s_{\ell}^{'}=\begin{cases}
s_{\ell} & \ell:\text{even}\\
-s_{\ell} & \ell:\text{odd}
\end{cases}$ and $s_{m}^{'}=\begin{cases}
s_{m} & m:\text{even}\\
-s_{m} & m:\text{odd}
\end{cases}$.

\begin{table*}
\begin{centering}
\begin{tabular}{ccccc|ccccc}
$\ell,m$ & $g$ & $\alpha$ & $s_{\ell}$ & $s_{m}$ & $\ell,m$ & $g$ & $\alpha$ & $s_{\ell}$ & $s_{m}$\tabularnewline
\hline 
$1,3$ & $\sqrt{\psi}$ & $\arccos\left(\frac{g^{-3}}{2}\right)$ & $+$ & $+$ & $1,8$ & $1.03945070$ & $1.46320427$ & $-$ & $-$\tabularnewline
$2,3$ & $\sqrt{\rho}$ & $\arccos\left(\frac{g^{3}}{2}\right)$ & $-$ & $+$ &  & $1.06621540$ & $1.25975111$ & $+$ & $+$\tabularnewline
$1,4$ & $1.06333694$ & $1.33527844$ & $-$ & $-$ &  & $1.08374370$ & $0.84015125$ & $+$ & $-$\tabularnewline
 & $1.18375182$$^{\dagger}$ & $0.90934345$ & $+$ & $+$ & $3,8$ & $1.01792429$ & $1.69767863$ & $-$ & $-$\tabularnewline
$3,4$ & $1.18375182$$^{\dagger}$ & $0.53405772$ & $-$ & $+$ &  & $1.06244389$ & $0.49612812$ & $+$ & $+$\tabularnewline
$1,5$ & $1.09900032$ & $1.73645968$ & $-$ & $+$ &  & $1.09231550$ & $0.71393754$ & $-$ & $-$\tabularnewline
 & $\sqrt{\rho}$ & $\arccos\left(\frac{g^{3}}{2}\right)$ & $+$ & $+$ &  & $1.10929363$ & $1.21438451$ & $-$ & $+$\tabularnewline
$2,5$ & $1.08646367$ & $0.80694026$ & $+$ & $+$ & $5,8$ & $1.03950336$ & $0.26297678$ & $-$ & $+$\tabularnewline
 & $1.16798953$ & $1.16141175$ & $-$ & $-$ &  & $1.09658675$ & $0.88770503$ & $-$ & $-$\tabularnewline
$3,5$ & $1.05036656$ & $0.42007091$ & $-$ & $+$ &  & $1.13377435$ & $1.12333647$ & $+$ & $+$\tabularnewline
 & $1.18711214$ & $1.38623505$ & $-$ & $-$ & $7,8$ & $1.06295569$ & $0.64055127$ & $+$ & $-$\tabularnewline
$4,5$ & $1.18738019$ & $0.43181263$ & $-$ & $+$ &  & $1.16615357$ & $0.27659675$ & $-$ & $+$\tabularnewline
$1,6$ & $1.04984644$ & $1.42286906$ & $-$ & $+$ & $1,9$ & $1.02209200$ & $1.29189202$ & $-$ & $-$\tabularnewline
 & $1.09917491$ & $1.14794978$ & $+$ & $-$ &  & $1.04695854$ & $1.66073000$ & $-$ & $+$\tabularnewline
 & $1.12611265$ & $0.57438369$ & $+$ & $+$ &  & $1.06444465$ & $1.11107685$ & $+$ & $+$\tabularnewline
$5,6$ & $1.03282504$ & $0.86317030$ & $+$ & $-$ &  & $1.07613313$ & $0.74087364$ & $+$ & $-$\tabularnewline
 & $1.18224537$ & $0.36320601$ & $-$ & $+$ & $2,9$ & $1.01283840$ & $0.58665015$ & $-$ & $-$\tabularnewline
$1,7$ & $1.01960526$ & $1.20613634$ & $-$ & $+$ &  & $1.04380602$ & $0.79080898$ & $+$ & $-$\tabularnewline
 & $1.06387323$ & $1.45420091$ & $+$ & $+$ &  & $1.06554885$ & $0.97639366$ & $-$ & $+$\tabularnewline
 & $1.09195331$ & $0.97020783$ & $+$ & $-$ &  & $1.08001175$ & $0.39672051$ & $+$ & $+$\tabularnewline
 & $1.10769105$ & $0.48526744$ & $+$ & $+$ & $4,9$ & $1.02868986$ & $0.45935343$ & $+$ & $+$\tabularnewline
$2,7$ & $1.05832758$ & $0.77578744$ & $-$ & $-$ &  & $1.06421568$ & $1.13693694$ & $-$ & $+$\tabularnewline
 & $1.09594733$ & $0.53256457$ & $+$ & $+$ &  & $1.08867435$ & $1.33411185$ & $+$ & $-$\tabularnewline
 & $1.11696283$ & $1.30397985$ & $-$ & $+$ &  & $1.10276124$ & $0.66651527$ & $-$ & $-$\tabularnewline
$3,7$ & $1.04634171$ & $0.58605974$ & $+$ & $+$ & $5,9$ & $1.01511363$ & $0.23291213$ & $-$ & $+$\tabularnewline
 & $1.09867941$ & $1.44528037$ & $-$ & $+$ &  & $1.05910448$ & $0.94513949$ & $+$ & $+$\tabularnewline
 & $1.12854879$ & $0.84668921$ & $-$ & $-$ &  & $1.09277920$ & $1.67194846$ & $-$ & $+$\tabularnewline
$4,7$ & $1.02518774$ & $0.29962941$ & $-$ & $+$ &  & $1.11272153$ & $0.73604039$ & $-$ & $-$\tabularnewline
 & $1.09707453$ & $1.22673682$ & $+$ & $+$ & $7,9$ & $1.03085468$ & $1.16917138$ & $+$ & $+$\tabularnewline
 & $1.14333477$ & $0.96167330$ & $-$ & $-$ &  & $1.08966388$ & $0.23866415$ & $-$ & $+$\tabularnewline
$5,7$ & $1.08331646$ & $0.30342198$ & $-$ & $+$ &  & $1.14226818$ & $1.46119977$ & $-$ & $-$\tabularnewline
 & $1.16177283$ & $1.43362675$ & $+$ & $+$ & $8,9$ & $1.01340552$ & $0.92556775$ & $-$ & $+$\tabularnewline
$6,7$ & $1.05175240$ & $0.73504742$ & $+$ & $-$ &  & $1.06962466$ & $0.5679008$ & $+$ & $-$\tabularnewline
 & $1.17446465$ & $0.31385868$ & $-$ & $+$ &  & $1.15808690$ & $0.24742995$ & $-$ & $+$\tabularnewline
\multicolumn{10}{c}{$^{\dagger}$$1.18375182=\sqrt{1.40126837}$ is the square root of
the smallest Salem number of degree $6$.}\tabularnewline
\end{tabular}
\par\end{centering}
\caption{\label{tab:lm}Table of all spiral chains up to $m=9$, corresponding
to different interaction distances. Note that $\left\{ \ell,m\right\} =\left\{ 2,3\right\} $
and $\left\{ \ell,m\right\} =\left\{ 1,5\right\} $ have exactly the
same $g$ and $\alpha$ and therefore can be combined in a single
spiral chain model.}
\end{table*}

\subsection{Power law steady state solutions\label{subsec:soln1}}

Substituting $\Phi_{n}\rightarrow Ak_{n}^{\alpha}$ in (\ref{eq:goy_gen}),
the nonlinear term vanishes when:
\begin{align*}
g^{\left(\alpha+3\right)m+\left(\alpha+1\right)\ell}-g^{m\left(\alpha+1\right)+\left(\alpha+3\right)\ell}\\
+g^{\left(\alpha+1\right)m-\left(2\alpha+3\right)\ell}-g^{\left(\alpha+3\right)m-\left(2\alpha+3\right)\ell}\\
+g^{\left(\alpha+3\right)\ell-\left(2\alpha+3\right)m}-g^{\left(\alpha+1\right)\ell-\left(2\alpha+3\right)m} & =0
\end{align*}
which can be satisfied if a) $\left(\alpha+1\right)=-\left(2\alpha+3\right)$
(i.e. $\alpha=-4/3$) independent of the value of $\ell$ and $m$,
in which case the first term cancels the fourth one, the second term
cancels the fifth and the third term cancels last one, or $b)$ $\left(\alpha+3\right)=-\left(2\alpha+3\right)$
(i.e. $\alpha=-2$), where the first term cancels the last one, second
term cancels the third one and the fourth term cancels the fifth one.
These correspond to the usual Kraichnan-Kolmogorov spectra $E\left(k\right)\propto\left\{ k^{-3},k^{-5/3}\right\} $
since $E\left(k_{n}\right)\equiv\Phi_{n}^{2}k_{n}$\citep{kraichnan:67}.
Note that these self-similar power law solutions on any spiral chain
$\mathcal{C}_{\ell,m}^{s_{\ell},s_{m}}$, may be anisotropic in the
sense that $\Phi_{k_{x},0}\neq\Phi_{0,k_{y}}$ for a given scale,
are isotropic in the sense that if we average over a few consecutive
scales we get a solution that is independent of the direction of $k$.

However, numerical integration of the model with energy injected roughly
in the middle of the spiral does not seem to converge to these solutions
(see sections \ref{eq:En_model} and \ref{sec:Numerical-Results}).
Instead it seems that the $\Phi_{n}$ act as ``random'' variables
and the system goes to a chain equipartition solution expected from
statistical equilibrium such that $P\left(\Phi_{n}\right)=e^{-\left(\beta_{1}k_{n}^{4}\left|\Phi_{n}\right|^{2}+\beta_{2}k_{n}^{2}\left|\Phi_{n}\right|^{2}\right)/2}$,
which gives (i.e. $T_{1}=\beta_{1}^{-1}$ and $T_{2}=\beta_{2}^{-1}$):
\[
\left\langle \left|\Phi_{n}\right|^{2}\right\rangle =\frac{T_{1}}{k_{n}^{4}+\frac{T_{1}}{T_{2}}k_{n}^{2}}
\]
and thus a spectral energy density scaling of the form $E\left(k\right)\propto\left\{ k^{-3},k^{-1}\right\} $
. In general which of these solutions will be observed depends on
various factors from numerical details to the way the system is driven.

\subsection{Energy and Enstrophy\label{subsec:Energy-and-Enstrophy:}}

Multiplying (\ref{eq:goy_gen}) by $\Phi_{n}^{*}k_{n}^{2}$ and taking
the real part, we can write the evolution of energy:

\begin{align}
\partial_{t}E_{n} & =\bigg[\left(g^{2m}-g^{2\ell}\right)t_{n+\ell}^{E}+\left(1-g^{2m}\right)t_{n}^{E}\nonumber \\
 & +\left(g^{2\ell}-1\right)t_{n-m+\ell}^{E}\bigg]+P_{n}^{E}-D_{n}^{E}\label{eq:En}
\end{align}
where $E_{n}=k_{n}^{2}\left|\Phi_{n}\right|^{2}$
\begin{equation}
t_{n}^{E}\equiv\text{Re}\left[g^{m-3\ell}k_{n}^{4}\sin\alpha_{qp}\Phi_{n-\ell+m}^{*}\Phi_{n-\ell}^{*}\Phi_{n}^{*}\right]\label{eq:tnE}
\end{equation}
or multiplying (\ref{eq:goy_gen}) by $\Phi_{n}^{*}k_{n}^{4}$,
\begin{align}
\partial_{t}W_{n} & =\bigg[\left(g^{2\left(m-\ell\right)}-1\right)t_{n+\ell}^{W}+\left(1-g^{2m}\right)t_{n}^{W}\nonumber \\
 & +\left(g^{2m}-g^{2\left(m-\ell\right)}\right)t_{n-m+\ell}^{W}\bigg]+P_{n}^{W}-D_{n}^{W}\label{eq:Wn}
\end{align}
where $W_{n}=k_{n}^{4}\left|\Phi_{n}\right|^{2}$, and
\begin{equation}
t_{n}^{W}\equiv\text{Re}\left[g^{m-3\ell}k_{n}^{6}\sin\alpha_{qp}\Phi_{n-\ell+m}^{*}\Phi_{n-\ell}^{*}\Phi_{n}^{*}\right]\;\text{.}\label{eq:tnW}
\end{equation}
It is easy to see that total energy $E=\sum_{n}E_{n}$ and total enstrophy
$W=\sum_{n}W_{n}$ are conserved since $t_{n}$'s cancel each-other
at different orders. This is basically due to the fact that each triad
conserves energy and enstrophy, and thus each chain of triads represented
by the spiral chain conserves energy and enstrophy independently.
Considering mid scale, well localized drive (say around the wave-number
$k_{f}$), with both large scale and small scale dissipations. If
we sum over (\ref{eq:En}) from $n=0$ up to an $n$ such that $k_{n}<k_{f}$,
in the inertial range for energy, we get:
\[
\partial_{t}\sum_{n'=0}^{n}E_{n'}+\Pi_{n}^{E}=-\varepsilon_{\ell}
\]
where $\varepsilon_{\ell}$ is the total large scale energy dissipation
and 
\begin{align*}
\Pi_{n}^{E} & \equiv-\bigg[\left(g^{2m}-g^{2\ell}\right)\sum_{j=1}^{m}t_{n-m+\ell+j}^{E}\\
 & +\left(1-g^{2m}\right)\sum_{j=1}^{m-\ell}t_{n-m+\ell+j}^{E}\bigg]
\end{align*}
A statistical steady state may imply:
\[
\overline{\Pi}_{n}^{E}=-\varepsilon_{\ell}
\]
and if $\overline{t}_{n}^{E}$ is independent of $n$ for an inertial
range, we can write
\[
\overline{\Pi}_{n}^{E}=-\lambda_{E}\overline{t}_{n}^{E}
\]
where $\lambda_{E}=\bigg[\left(1-g^{2\ell}\right)m-\left(1-g^{2m}\right)\ell\bigg]$.
Note that for $g=1+\epsilon$, so that $g^{2\ell}=1+2\ell\epsilon+\left(2\ell^{2}-\ell\right)\epsilon^{2}$
and finally $\lambda=2\left(m-\ell\right)m\ell\epsilon^{2}>0$, since
$m>\ell$. If we increase $g$, $\lambda>0$ will be more easily satisfied.
So practically for any $g>1$ and $\ell>m$, we have $\lambda>1$.

Note that the instability assumption of a single triad discussed in
Section \ref{sec:single_triad} for an abritrary triad implies $\overline{t}_{n}>0$,
resulting in an inverse cascade of energy (i.e. $\overline{\Pi}_{n}^{E}<0$).
Similarly by computing the sum over (\ref{eq:Wn}) from $n$ to $N$
such that $k_{n}>k_{f}$ is in the inertial range for enstrophy:
\[
\partial_{t}\sum_{n'=0}^{n}W_{n'}-\Pi_{n}^{W}=-\varepsilon_{s}
\]
where $\varepsilon_{s}$ is the total small scale dissipation and
\begin{align*}
\Pi_{n}^{W} & \equiv\bigg[\left(g^{2\left(m-\ell\right)}-1\right)\sum_{j=1}^{m}t_{n-m+\ell+j}^{W}\\
 & +\left(1-g^{2m}\right)\sum_{j=1}^{m-\ell}t_{n-m+\ell+j}^{W}\bigg]
\end{align*}
is the $k$-space flux of enstrophy. Assuming that in the inertial
range $\overline{t}_{n}^{W}$ remain independent of $n$, we get:
\[
\overline{\Pi}_{n}^{W}=\lambda_{W}\overline{t}_{n}^{W}
\]
where
\[
\lambda_{W}\equiv\left(1-g^{2m}\right)\left(m-\ell\right)-\left(1-g^{2\left(m-\ell\right)}\right)m>0
\]
which can be seen from the fact that $\lambda_{W}$ has the same form
as $\lambda_{E}$ but $\ell$ replaced by $m-\ell$, and $m-\ell<m$.
The instability assumption for a single triad suggests $\overline{t}_{n}^{W}>0$,
so we get a forward cascade of enstrophy.

\subsection{Zero flux solutions\label{subsec:Zero-flux-solutions}}

A zero flux solution for the energy can be obtained more easily for
specific values of $\ell$ and $m$. We therefore consider the case
$\ell=1$, $m=3$ first, with $\Pi_{n}^{E}=0$, which gives
\begin{align*}
\overline{\Pi}_{n}^{E} & \equiv-\bigg[\left(g^{6}-g^{2}\right)\left(\overline{t}_{n-1}^{E}+\overline{t}_{n}^{E}+\overline{t}_{n+1}^{E}\right)\\
 & +\left(1-g^{6}\right)\left(\overline{t}_{n-1}^{E}+\overline{t}_{n}^{E}\right)\bigg]=0\quad\text{.}
\end{align*}
Assuming $\overline{t}_{n+1}=g^{\mu}\overline{t}_{n}$ we get:
\[
g^{2\mu}g^{2}\left(g^{2}+1\right)-g^{\mu}-1=0\;\text{,}
\]
whose solution is $\mu=\ln\bigg[\frac{1}{2g^{2}\left(g^{2}+1\right)}\pm\frac{1}{2g^{2}\left(g^{2}+1\right)}\sqrt{1+4g^{2}\left(g^{2}+1\right)}\bigg]/\ln g=-2$,
for $g=\sqrt{\psi}$. Since $\overline{t}_{n}\sim k_{n}^{-2}$, and
$\overline{t}_{n}\propto k_{n}^{4}\left|\Phi_{n}\right|^{3}$ , one
obtains a spectral energy density of the form $E\left(k_{n}\right)=\left|\Phi_{n}\right|^{2}k_{n}\propto k_{n}^{-3}$.
Similarly, the zero enstrophy flux solution for $\ell=1$, $m=3$
gives $\mu=2$, which means $\overline{t}_{n}^{W}\sim k_{n}^{2}$
and therefore $E\left(k_{n}\right)\propto k_{n}^{-5/3}$. The fact
that \emph{the zero flux solution for energy gives the same scaling
as the forward enstrophy cascade solution (i.e. $E\left(k\right)\propto k^{-3}$)
and the zero flux solution for enstrophy gives the same scaling as
the inverse energy cascade solution (i.e. $E\left(k\right)\propto k^{-5/3}$)},
is a nice feature of the spiral chain structure.

In order to see if this works in the general case, we can substitue
$\overline{t}_{n}\propto k_{n}^{-2}$ into the general expressions
for the energy flux:
\begin{align*}
\Pi_{n}^{E} & \equiv-\bigg[\left(g^{2m}-g^{2\ell}\right)\sum_{j=0}^{m-1}g^{j\mu}\\
 & +\left(1-g^{2m}\right)\sum_{j=0}^{m-\ell-1}g^{j\mu}\bigg]t_{n-m+\ell+j}^{E}=0\;\text{.}
\end{align*}
Using the relation 
\[
\sum_{j=0}^{m-1}g^{j\mu}=\frac{\left(1-g^{m\mu}\right)}{\left(1-g^{\mu}\right)}
\]
one can see that the energy flux vanishes if
\[
\bigg[\left(g^{2m}-g^{2\ell}\right)\frac{\left(1-g^{m\mu}\right)}{\left(1-g^{\mu}\right)}+\left(1-g^{2m}\right)\frac{\left(1-g^{\left(m-\ell\right)\mu}\right)}{\left(1-g^{\mu}\right)}\bigg]=0\;\text{.}
\]
We find that $\mu=-2$ is a solution of this, since if we substitute
it into the above expression, we get
\[
\bigg[\left(g^{2m}-g^{2\ell}\right)\left(1-g^{-2m}\right)+\left(1-g^{2m}\right)\left(1-g^{-2\left(m-\ell\right)}\right)\bigg]=0
\]
This means that for any combination of $\ell$, $m$ and $g$, $\overline{t}_{n}^{E}\propto k_{n}^{-2}$,
gives a zero flux solution of energy with $E\left(k\right)\propto k^{-3}$.

Similarly it is easy to see that $\mu=2$ is a solution of the general
relation for the vanishing enstrophy flux
\begin{align*}
\Pi_{n}^{W} & \equiv\bigg[\left(g^{2\left(m-\ell\right)}-1\right)\frac{\left(1-g^{m\mu}\right)}{\left(1-g^{\mu}\right)}\\
 & +\left(1-g^{2m}\right)\frac{\left(1-g^{\left(m-\ell\right)\mu}\right)}{\left(1-g^{\mu}\right)}\bigg]t_{n-m+\ell+j}^{W}=0
\end{align*}
resulting in $\overline{t}_{n}^{W}\propto k_{n}^{2}$ and therefore
$E\left(k\right)\propto k^{-5/3}$.

\section{The model for $E_{n}$\label{sec:The-model-for}}

The general model for the evolution of turbulent energy on the spiral
chain can be formulated as

\begin{align}
\partial_{t}E_{n} & =\bigg[\left(g^{2m}-g^{2\ell}\right)t_{n+\ell}^{E}+\left(1-g^{2m}\right)t_{n}^{E}\nonumber \\
 & +\left(g^{2\ell}-1\right)t_{n-m+\ell}^{E}\bigg]+P_{n}^{E}-D_{n}^{E}\label{eq:En_model}
\end{align}
where
\begin{equation}
t_{n}^{E}=g^{-\ell}k_{n}\sin\left[\left(m-\ell\right)\alpha\right]E_{n}^{3/2}\label{eq:tn2}
\end{equation}
Note that $E\left(k_{n}\right)=E_{n}k_{n}^{-1}$ and that $E_{n}>0$
and $P_{n}^{E}>0$ to assure realizability. The model still conserves
energy and enstrophy, and results in a clean dual cascade solution.
And the difference from a model that solves the complex amplitudes
$\Phi_{n}$ is mainly in the definition (\ref{eq:tnE}) vs. (\ref{eq:tn2}).
The two models would become ``equivalent'' if the sums of the complex
phases would vanish at each scale (for example for $\ell=2$, $m=3$,
this would mean $\phi_{n}+\phi_{n+1}-\phi_{n-2}=0$, where $\phi_{n}$
are the complex phases). The condition is nontrivial and is not satisfied
in the nonlinear stage by a complex chain model for $\Phi_{n}$. Hence
the complex chain fails to describe the cascade but instead evolves
towards statistical chain equipartition.

Model in (\ref{eq:En_model}), works for any $\ell$ and $m$ combination
given in Table \ref{tab:lm}, but one should pay attention to the
fact that as $\ell$ and $m$ change, $g$, and therefore the range
of wave-numbers that are covered by the model changes, which means
that the dissipation and the boundary terms should also be modified
accordingly. Note finally that the assumption of $t_{n}^{E}\propto k_{n}E_{n}^{3/2}$
corresponds to the Kovasznay's form.

\subsection{Continuum limit\label{subsec:Continuum-limit}}

It is also possible to interpret (\ref{eq:En_model}) as a shell model
by disregarding the information on angles and therefore lifting the
restriction on $g$ values. In this case the resulting model is a
simple discrete formulation of a general model where any value of
$g$ is allowed and an arbitrary factor {[}instead of the $\sin\left(m-\ell\right)\alpha${]}
multiplies the nonlinear term, as in shell models. This interpretation
allows us to transform the problem into a differential approximation
model by considering the continuum limit of (\ref{eq:En_model}),
with $\ell=1$, $m=2$, by considering $g\rightarrow1+\epsilon$.
Defining $E\left(k\right)=E_{n}k_{n}^{-1}$ and $F\left(k\right)=k^{3/2}E\left(k\right)^{3/2}$,
so that $k_{n+1}=k\left(1+\epsilon\right)$ and $k_{n-1}\left(1-\epsilon+\epsilon^{2}\right)$
so that
\begin{figure}
\begin{centering}
\includegraphics[width=0.99\columnwidth]{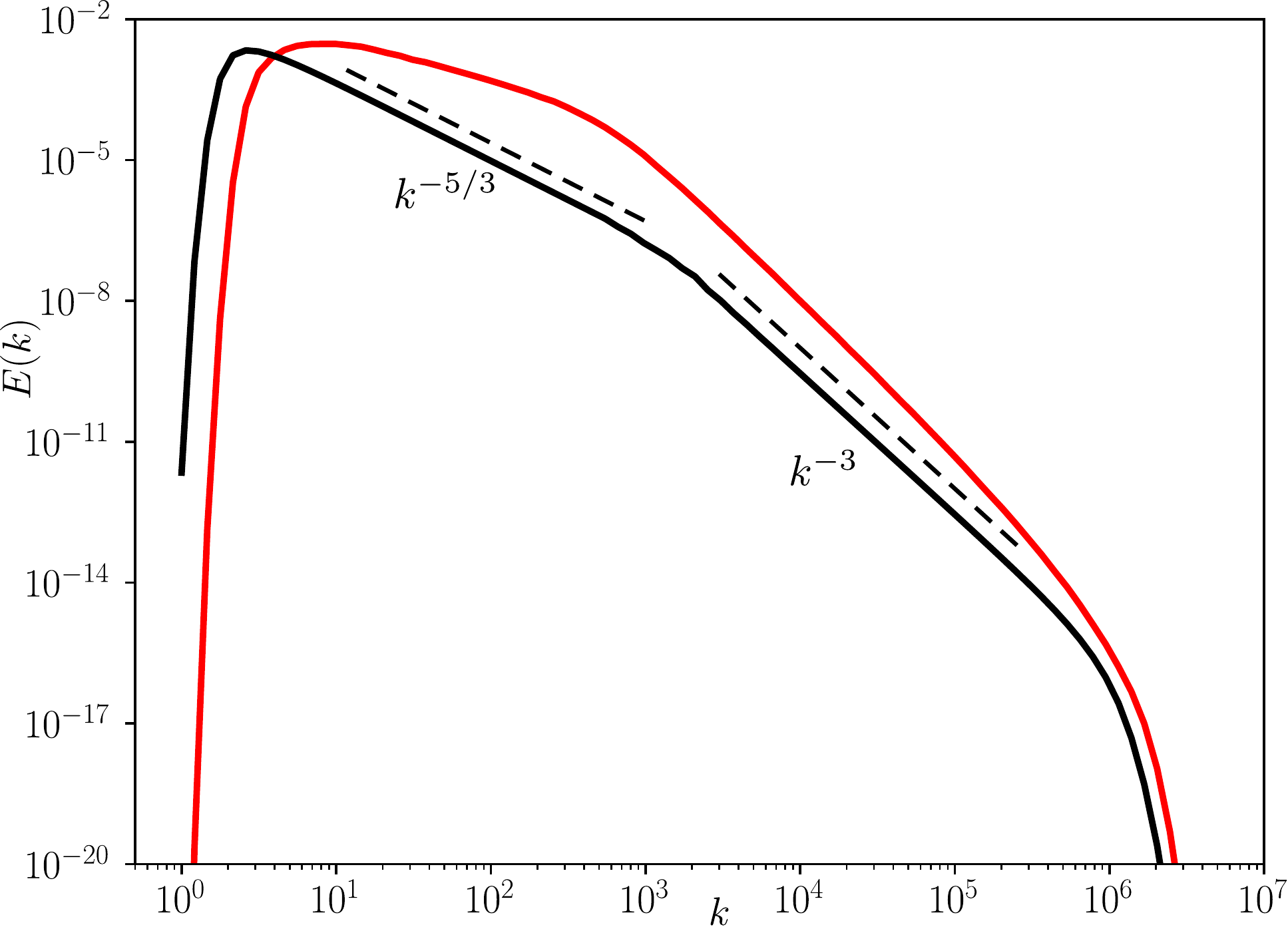}
\par\end{centering}
\caption{\label{fig:spec_scm13}wave-number spectra for the two variants of
the $\ell=1$, $m=3$ spiral chain model. The red line is the model
for the complex amplitude $\Phi_{n}$, whereas the black line is the
model for $E_{n}$. While the model for $E_{n}$ is driven with constant
forcing $P_{n}=2.5\times10^{-4}$, the model for $\Phi_{n}$ is driven
with random forcing such that $\left\langle P_{n}\right\rangle =2.5\times10^{-4}$.
The spectrum for the $\Phi_{n}$ model is averaged over a long stationary
phase, where $E\left(k_{n}\right)=\left\langle \left|\Phi_{n}\right|^{2}\right\rangle k_{n}$,
which is integrated up to $t=10000$ and the average is computed over
$t=\left[5000,10000\right]$, whereas the spectrum for the $E_{n}$
model is averaged over $t=\left[190,200\right]$ (in fact the instanteneous
solution is not that different from the averaged result).}
\end{figure}
\[
F\left(k_{n+1}\right)\approx\left(F+k\epsilon\frac{dF}{dk}+\frac{1}{2}\epsilon^{2}k^{2}\frac{d^{2}F}{dk^{2}}\right)
\]
\[
F\left(k_{n-1}\right)\approx\left(F\left(k\right)-k\left(\epsilon-\epsilon^{2}\right)\frac{dF}{dk}+\frac{1}{2}\epsilon^{2}k^{2}\frac{d^{2}F}{dk^{2}}\right)
\]
and
\begin{align*}
 & \bigg[g^{2}k_{n}^{-1}t_{n+1}^{E}-\left(1+g^{2}\right)k_{n}^{-1}t_{n}^{E}+k_{n}^{-1}t_{n-1}^{E}\bigg]\\
 & \approx3\epsilon^{2}F+5\epsilon^{2}k\frac{dF}{dk}+\epsilon^{2}k^{2}\frac{d^{2}F}{dk^{2}}\;\text{.}
\end{align*}
This finally gives:
\begin{equation}
\partial_{t}E-C\frac{\partial}{\partial k}\left(k^{-1}\frac{\partial}{\partial k}\left(k^{9/2}E^{3/2}\right)\right)=P_{E}\left(k\right)-D_{E}\left(k\right)\label{eq:dam}
\end{equation}
as a differential approximation model\citep{leith:67}. It is clear
that the two solutions $E\left(k\right)\propto k^{-5/3}$ and $E\left(k\right)\propto k^{-3}$
both cause the nonlinear term to vanish. In fact the way the flux
is approximated, it works nicely that $k^{-5/3}$ gives a constant
and negative energy flux. In fact the constant flux solution of the
above equation is E$\left(k\right)=\left(\frac{\varepsilon_{\ell}}{2C}\right)^{2/3}k^{-5/3}$,
which is helpful for picking the value of $C$ in order to normalize
the model properly. The continuum limit as discussed above results
in an isotropic model, since its derivation starts from a shell-model
with no regards to angles. 
\begin{figure}
\begin{centering}
\includegraphics[width=0.99\columnwidth]{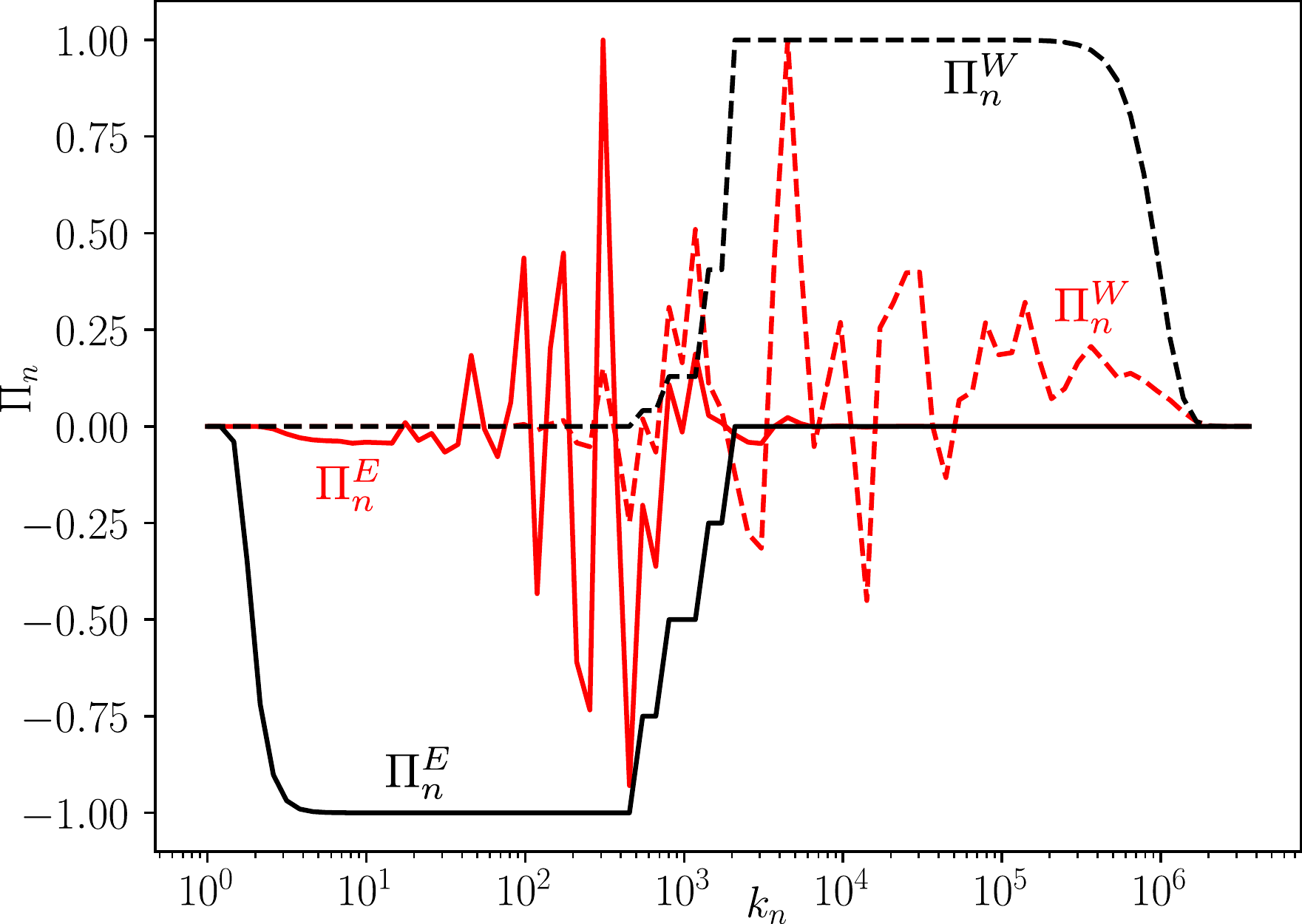}
\par\end{centering}
\caption{\label{fig:flux_scm13}Energy and enstrophy fluxes for the two variants
of the $\ell=1$, $m=3$ spiral chain model. The red solid line and
the red dashed line are the energy and enstrophy fluxes for the complex
amplitude model, whereas the black solind line and the black dashed
line are the energy and enstrophy fluxes for the $E_{n}$ model respectively,
normalized to their maximum values. We can see that rapid oscillations
of the phases observed in the complex model causes the suppression
of the fluxes and results in statistical chain equipartition solutions
instead of proper dual cascade solutions.}
\end{figure}

\subsection{4-Spiral Chain Model\label{subsec:4-Spiral-Chain-Model}}

Considering the model in (\ref{eq:spi_full_pre}) and using 4 such
spiral chains that are basically rotated by $\delta\alpha=j\alpha/4$
and scaled by $g^{j/4}$ where $j=1,2,3$ with respect to the original
spiral (together with the original spiral itself, see fig. \ref{fig:grid}
) gives us a 4-spiral chain model, where the each spiral chain is
coupled with itself but not with the other three. The advantage of
the existence of the other chains is therefore a better coverage of
the k-space but not a better description of the nonlinear interaction
(i.e. the number of triads in the 4 spiral chain model is basically
4 times the single spiral chain one). Such a model can be formulated
alternatively by defining $g=\rho^{1/8}$ and $\alpha=\frac{1}{4}\arccos\left(-\frac{g^{12}}{2}\right)$
and using \textbf{$\mathbf{k}_{n}=k_{n}\left(\cos\alpha_{n},\sin\alpha_{n}\right)$},
where $k_{n}=k_{0}g^{n}$ and $\alpha_{n}=\alpha n$ as usual (note
that $g$ here is obviously different from the earlier one). The evolution
for $\Phi$ can then be written as
\begin{align}
\partial_{t}E_{n} & =k_{n}\sin\alpha\bigg[g^{16}\left(g^{8}-1\right)E_{n+8}^{3/2}+\left(g^{32}-1\right)g^{-12}E_{n+4}^{3/2}\nonumber \\
 & +\left[g^{-8}-2g^{16}+g^{-24}\right]E_{n}^{3/2}+\left(g^{16}-1\right)g^{-12}E_{n-4}^{3/2}\nonumber \\
 & +\left(g^{8}-1\right)g^{-40}E_{n-16}^{3/2}\bigg]+P_{n}^{E}-D_{n}^{E}\label{eq:fcm6E}
\end{align}
Please note the simplicity of the nonlinear couplings in this model.
Albeit the fact that the model considers two kinds of triangles and
spans roughly about 10 different directions for a given ``scale''
it represents these nonlinear interactions with only $5$ terms.

The spiral grid corresponding to the 4-spiral chain, and its reflection
with respect to the origin is also shown in figure \ref{fig:grid}.
The grid provides an alternative way of looking at the spiral chain
as a partition of the $k$-space. The surface element for a given
cell $n$, can then be written as:
\begin{figure}
\begin{centering}
\includegraphics[width=0.98\columnwidth]{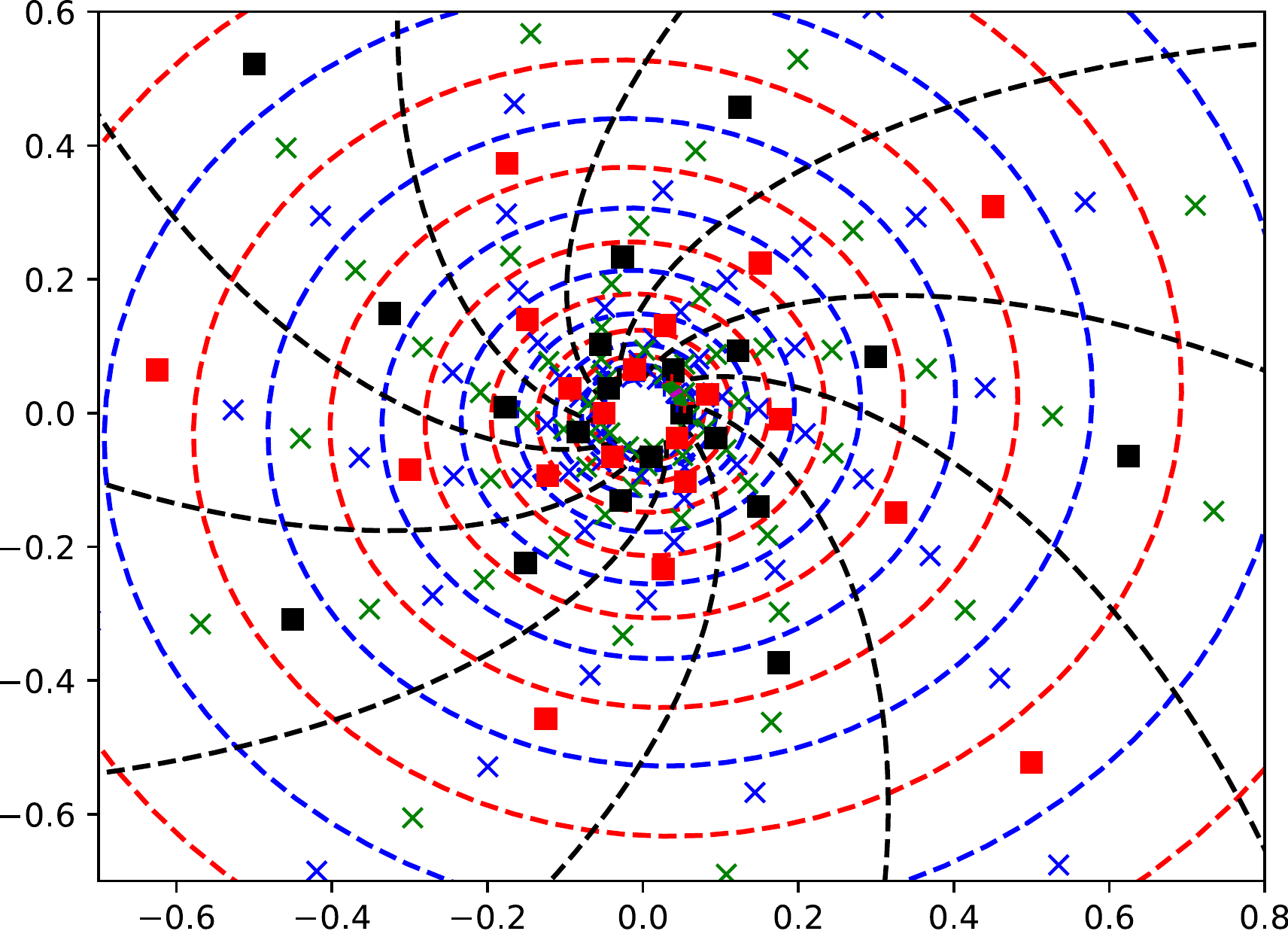}
\par\end{centering}
\caption{\label{fig:grid}The 4-spiral chain grid shown explicitly. The original
spiral chain is shown in black squares, while its reflection with
respect to the origin is shown in red squares. The full system is
symmetric with respect to reflection $\mathbf{k}\rightarrow-\mathbf{k}$,
and therefore one can actually use only half of the $k$-plane (e.g.
the upper half) and obtain the rest of the points by reflection.}
\end{figure}
\[
S_{n}=\frac{\pi\left(g^{1}-g^{-1}\right)\left(g^{5}-g^{-5}\right)}{20\ln\left(g\right)}k_{n}^{2}\approx0.03534\times\pi k_{n}^{2}
\]
which is basically a small percentage of the area of the circle with
that same radius. One obvious problem with this perspective is the
``hole'' that it leaves at the center. One can remedy this either
by computing the actual shape of the leftover region and adding it
as a partition cell, or alternatively adding a circular cell around
the origin and reducing the surface elements of the first few cells
of the partition by subtracting the part of the circular region that
intersects with the cell that is left for the circular element defined
at the origin. While rather promising, spiral partitioning of $k$-space
is not the focus of this paper. Thus we leave it for future studies
to resolve its particular issues.
\begin{figure}
\begin{centering}
\includegraphics[width=0.99\columnwidth]{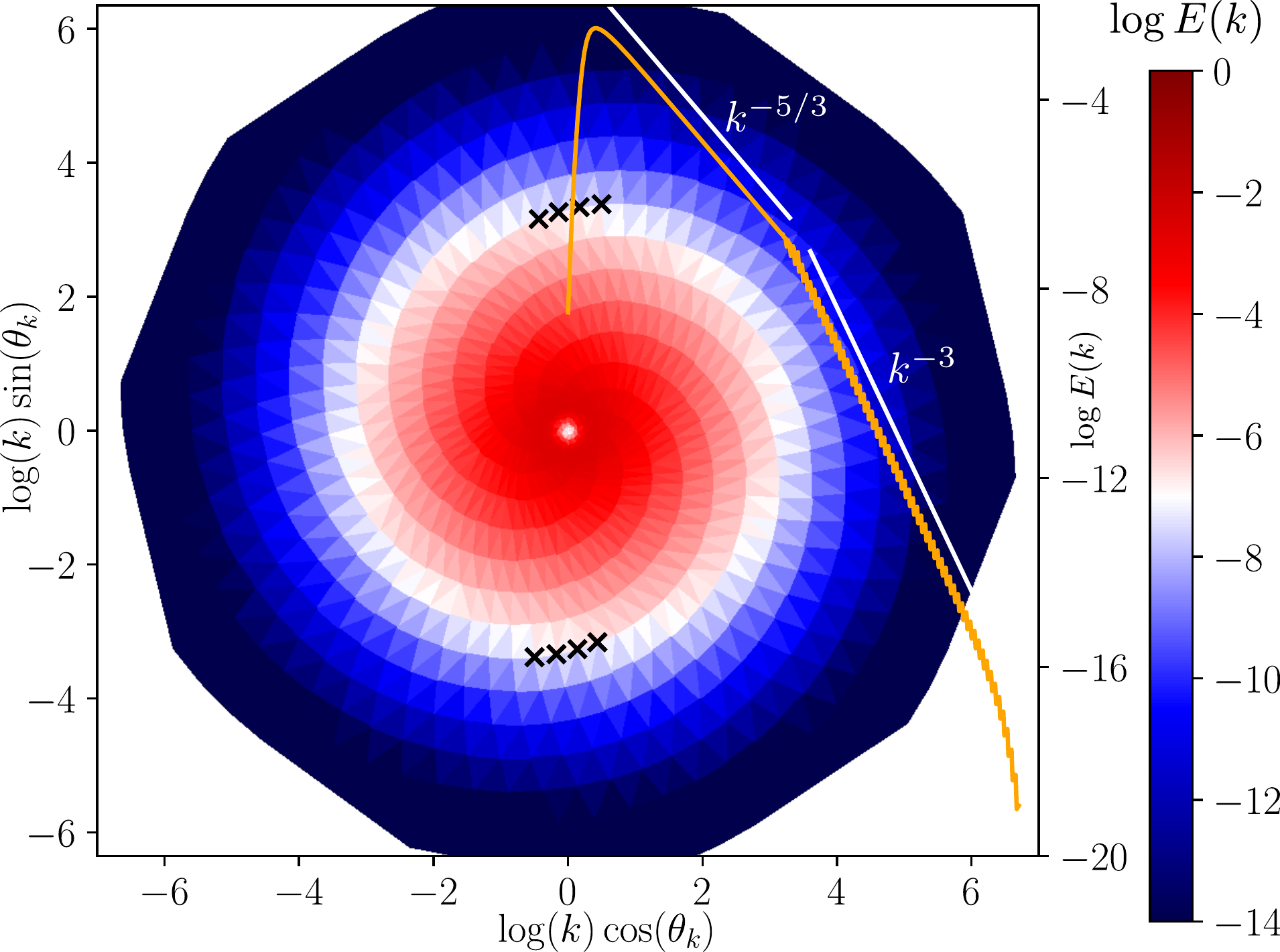}
\par\end{centering}
\caption{\label{fig:fcm2D}Two dimensional ``log-log'' {[}i.e. $\left\{ \log\left(k\right)\cos\left(\theta_{k}\right),\log\left(k\right)\sin\left(\theta_{k}\right),\log\left(E\left(k\right)\right)\right\} ${]}
plot of the wave-number spectrum for the 4-spiral chain model discussed
in Section \ref{subsec:4-Spiral-Chain-Model}. The energy injection
is located around $k_{x}=0$, $k_{y}=\pm2\times10^{3}$, shown above
as black $\times$'s. The resulting spectrum consists of a clear inverse
energy cascade range of $E\left(k\right)\propto k^{-5/3}$ (the red
region), and a forward enstrophy cascade range of $E\left(k\right)\propto k^{-3}$
(the blue region). One dimensional spectrum, which can be obtained
by plotting E$\left(k_{n}\right)=E_{n}/k_{n}$ as a function of $k_{n}=\left|\mathbf{k}_{n}\right|$,
is also shown with guiding lines showing the theoretical predictions.}
\end{figure}

\section{Numerical Results\label{sec:Numerical-Results}}

Existence of all possible triads enabled by neatly matching grid points
of a regular mesh allows important advantages such as good statistical
behavior, mathematical clarity and use of efficient numerical methods
such as fast fourier transforms. The models that we present in this
paper are not likely to replace direct numerical simulation schemes
such as pseudo-spectral methods even when very large wave-number ranges
are needed. Instead, they may be used as models of cascade that can
provide a mathematical framework for understanding the detailed structure
of the cascade process through self-similar triad interactions.

Various models introduced in this paper, can be considered as sets
of ordinary differential equations that can be solved numerically
in the presence of well-localized forcing and dissipation in the hope
of establishing numerical inertial range cascade behavior. However,
note that the primary goal of this paper is to introduce the framework
of spiral chains and not to perform a detailed numerical study of
these models.
\begin{figure}
\centering{}\includegraphics[width=0.99\columnwidth]{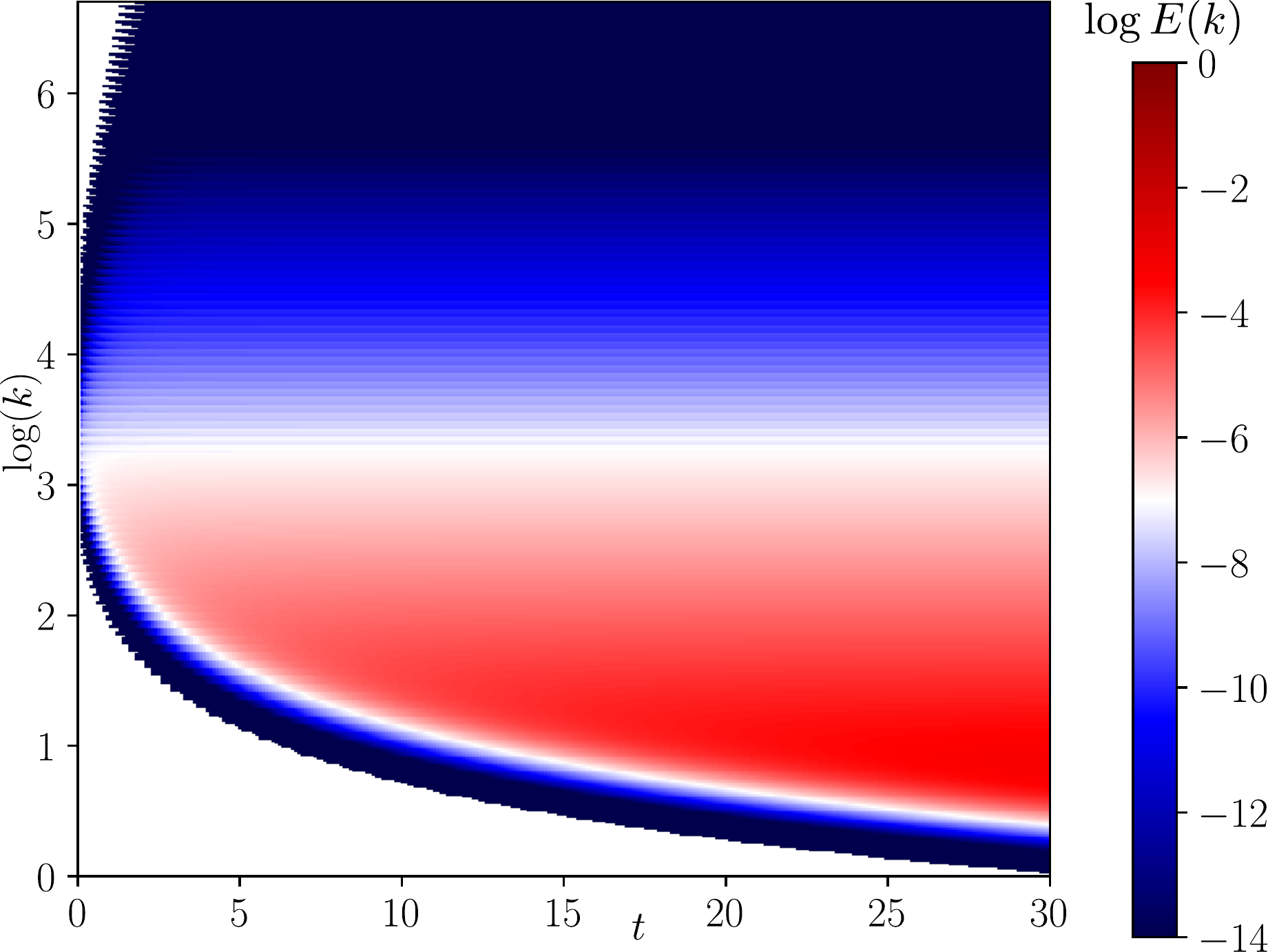}\caption{\label{fig:fcm_t}Time evolution of the one dimensional $k$-spectrum,
showing how it gets established in time in an asymmetric nonlinear
diffusion where the small scales are rapidly filled while large scales
take a while to populate. Here the colors show different levels of
$E\left(k\right)$, where the red region correspond to inverse cascade
and the blue region corresponds to forward cascade as in Figure \ref{fig:fcm2D}.}
\end{figure}

The results for the basic chain model for complex amplitudes $\Phi_{n}$'s
for the chain $\ell=1$, $m=3$, driven with stochastic forcing, with
dissipation of the form $D_{n}=\left(\nu k^{4}+\nu_{L}/k^{6}\right)\Phi_{k}$,
can be seen in figure \ref{fig:spec_scm13} and Figure \ref{fig:flux_scm13}
along with the model for $E_{n}$ for comparison. Even though the
evolution of the complex phase is due to nonlinear couplings, the
phases rapidly become ``random'' in practice, causing the fluxes
to oscillate (both in time and along the chain), resulting in a statistical
chain equipartition solution, which overwhelms the cascade process.
In contrast the results for the chain model for $E_{n}$ for $\ell=1$,
$m=3$ show a clear dual cascade and thus a distinct Kraichnan-Kolmogorov
spectrum. Here we used a simple python solver\citep{gurcan:github:19},
based on scipy ode solver\citep{scipy}.

\begin{figure}
\centering{}\includegraphics[width=0.99\columnwidth]{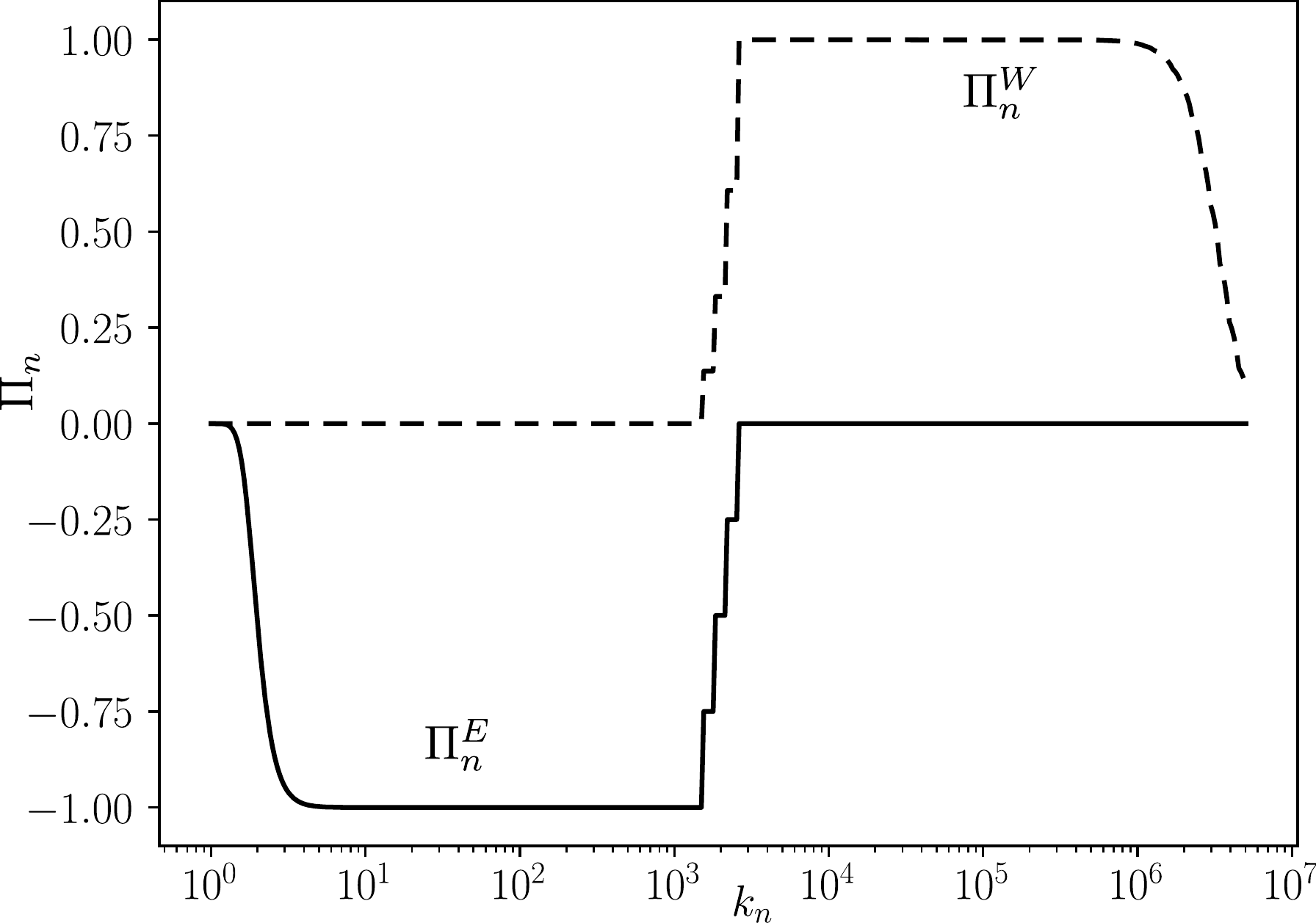}\caption{\label{fig:fluxes}Energy and enstrophy fluxes $\Pi_{n}^{E}$ and
$\Pi_{n}^{W}$, normalized to their maximum values, for the 4-spiral
chain model. This is averaged over 10 time steps, but even instantaneously,
they are extremely flat and stationary.}
\end{figure}

The four chain model introduced in section \ref{subsec:4-Spiral-Chain-Model}
has a good coverage of the $k$-space both in radial and in angular
directions. Here, we present the two dimensional wave number spectrum
that we obtain from this model, with $N=440$, $\nu=10^{-24}$, $\nu_{L}=10$,
and anisotropic forcing $P_{n}^{E}=2.5\times10^{-4}$ for the $4$
wave-numbers closest to $k_{x}=0$, $k_{y}=\pm2\times10^{3}$ in Figure
\ref{fig:fcm2D}. Even though the drive is anisotropic, the resulting
spectrum is isotropic since the flux along the spiral chain results
naturally in isotropization of the spectrum. The time evolution of
the wave-number spectrum is shown in Figure \ref{fig:fcm_t} and the
fluxes are shown in Figure \ref{fig:fluxes}. Finally no intermittency
has been observed in any of the models for $E_{n}$, since $S_{j}\left(k_{n}\right)\equiv\left\langle E_{n}^{j/2}\right\rangle \sim k_{n}^{-j/3}$
for the inverse cascade range and $S_{j}\left(k_{n}\right)\equiv\left\langle E_{n}^{j/2}\right\rangle \sim k_{n}^{-j}$
for the forward cascade range, with no discernible correction.

\section{Conclusion\label{sec:Conclusion}}

The geometry of the self-similar dual cascade in two dimensions as
the energy or enstrophy is transferred from one wave-vector to another
through triadic interactions are considered. The resulting picture
is that of a chain of triangles that are rotated and scaled, such
that the smallest wave-number of one triangle becomes the middle and
largest wave-numbers of the consecutive triads. A particular class
of triangles, make it such that one can form a regular logarithmic
spiral grid out of the wave-numbers $\mathbf{k}_{n}=k_{0}\left(ge^{i\alpha}\right)^{n}$,
where the complex number is interpreted as a two-dimensional vector
so that the real and imaginary parts are the $x$ and $y$ components,
with $g$ and $\alpha$ being the scaling factor and the divergence
angle respectively. Nonlinear interactions take place among the wave
vectors $\mathbf{k}_{n}$, $\mathbf{k}_{n+\ell}$ and $\mathbf{k}_{n+m}$
on such a spiral, where the values of $\ell$ and $m$ define (not
necessarily uniquely) particular values of $g$ and $\alpha$. There
is in fact a large number of such triangles, some of which are listed
explicitly in table \ref{tab:lm}. It is argued that the self-similar
cascade takes place along triad chains, and therefore the concept
of spiral chains can give us furhter insight into this mechanism,
without the explicit assumption of isotropy.

In order to demonstrate the usefullness of the concept, a series of
spiral chain models both for the complex amplitudes $\Phi_{n}$ as
well as energy $E_{n}$ have been developed. It is shown that analytical
solutions of these models with constant or zero flux cases agree with
the Kraichnan-Kolmogorov phenomonology of isotropic cascade. While
the complex models, that are basically ``shell models'' with elongated
triads can not numerically reproduce the dual cascade (because the
nonlinear evolution of the phases, lead to oscilattory solutions for
the fluxes of conserved quantities), and instead converge to unphysical
chain equipartition solutions. The model for $E_{n}$ in (\ref{eq:En_model})
can reproduce the dual cascade results numerically for any $\ell$
and $m$.

In particular, a 4-spiral chain model for $E_{n}$ is introduced in
(\ref{eq:fcm6E}), which has good angular coverage and has two kinds
of triads thanks to the choice of $g$ and $\alpha$ to include $\ell=2$,
$m=3$ and $\ell=1$, $m=5$ simultaneously. While a simple test of
anisotropic energy injection leads to the usual isotropic dual cascade
result, the model can be developed for self-consistent drive or other
similar cases for more complex problems such as two dimensional plasmas
or geophysical fluids.
\begin{acknowledgments}
The authors would like to thank P. H. Diamond, W-C. Müller and attendants
of the \emph{Festival de Théorie, Aix en Provence} in 2017.
\end{acknowledgments}

\end{document}